\documentclass{revtex4-2}

\usepackage{graphicx}
\usepackage{dcolumn}
\usepackage{bm}
\usepackage{mathtools,amssymb,lipsum}

\begin{document}

\title{Detection of Orbital Angular Momentum Modes via Spiral Phase Plate}

\author{Amin Hakimi$^1$}

\author{S. Faezeh Mousavi$^{2,3}$}

\author{Amir Nader Askarpour$^1$}
\email{askarpour@aut.ac.ir}
\affiliation{$^1$Department of Electrical Engineering, Amirkabir University of Technology, Tehran, Iran\\
$^2$Department of Physics, University of Trieste, Trieste, Italy
\\$^3$National Institute of Optics (CNR-INO), Area Science Park, Basovizza, Trieste, Italy}

\begin{abstract}
Mode division multiplexing (MDM) systems leveraging spatial modes carrying orbital angular momentum (OAM) present a promising approach to enhance communication capacity in free-space and fiber-optic networks. Efficient detection of OAM modes is critical for their practical implementation. Spiral phase plates (SPPs) are low-cost optical elements with simple structures, capable of generating OAM waves with high conversion efficiency and precision. This makes the inverse-SPP a promising candidate for the detection of OAM modes. In this study, the performance of SPPs as OAM detectors is thoroughly analyzed. Key parameters, including efficiency, crosstalk, and signal-to-interference ratio, are examined through analytical and numerical methods. Results demonstrate that an inverse-SPP, combined with an optimized propagation length and aperture size, enables effective and precise OAM detection operation.
\\
\\
\textit{keywords}---Orbital Angular Momentum, Spiral Phase Plate, Free-Space Optical Communication
\end{abstract}

\maketitle

\section{Introduction}
\label{sec: Int}
Conventional multiplexing techniques, including time-division multiplexing (TDM), wavelength-division multiplexing (WDM), and polarization-division multiplexing (PDM), have been pivotal in efficiently utilizing bandwidth by modulating independent channels across time, wavelength, and polarization, respectively \cite{tucker1988optical,brackett1990dense,ivanovich2018polarization}. While these techniques partially addressed the escalating demand for high transmission capacities in optical networks \cite{kobayashi201145,zhou201164,qian2012high,chvojka2020visible}, their scalability is inherently constrained by the Shannon limit and fiber nonlinearities \cite{mitra2001nonlinear}. To overcome these barriers, space-division multiplexing (SDM) has emerged as a promising paradigm. By leveraging the optimized cross-sectional area of optical fibers through multi-core or multi-mode designs, SDM provides an avenue for substantial capacity enhancements against the capacity crunch issue \cite{richardson2013space,puttnam2021space}. 

Among the SDM approaches, mode-division multiplexing (MDM) in various configurations of fibers (such as multi-mode, few-mode, multi-core, and coupled-core schemes) offers a compelling solution by exploiting orthogonal spatial modes as additional degrees of freedom. This allows independent data modulation within distinct bandwidth channels, effectively enhancing the utilization of the fiber's spatial properties \cite{berdague1982mode,ryf2012mode,sillard2014few,su2021perspective}. One of the recently interested higher-order spatial modes for communication systems is orbital angular momentum (OAM) characterized by its helical phase structure of $e^{i\ell \phi}$ ($\phi$ is the azimuthal angle and $\ell=0,\pm1, \pm2, \ldots$ denotes the topological charge). Each OAM mode 
carries angular momentum of $\ell \hbar$ per optical mode, providing a theoretically unbounded mode count\cite{yao2011orbital,willner2021orbital}. 
This property has been harnessed to significantly expand the traffic capacity of optical communication systems as demonstrated in both free-space \cite{wang2012terabit,wang2022orbital,singh2022performance} and fiber-based \cite{bozinovic2013terabit,wang2021orbital,liu20221} optical channels. 

\begin{figure}[t]
	\centering
	\includegraphics[width=0.5\linewidth]{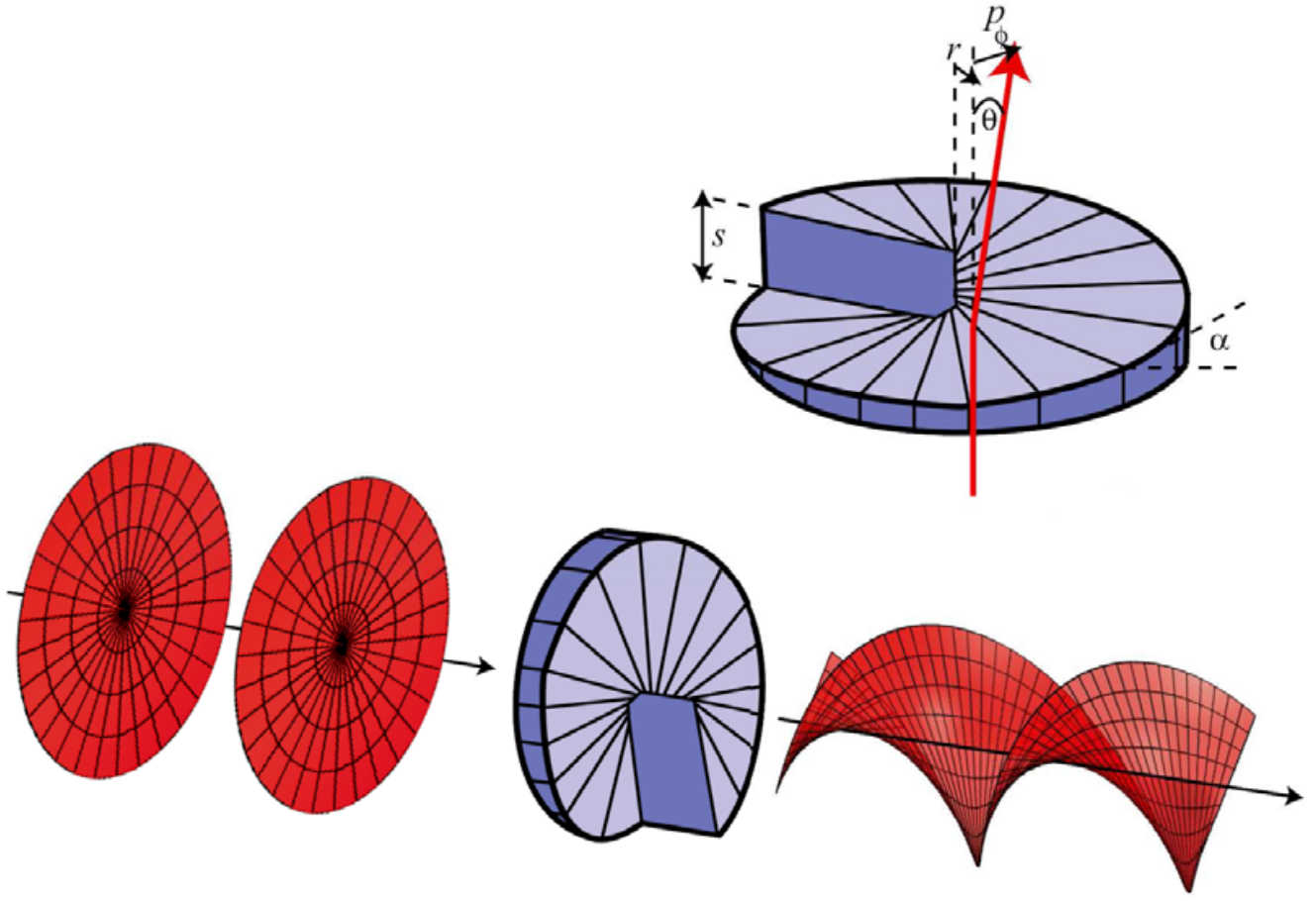}
	\caption{A spiral phase plate is used to convert a Gaussian beam into a helically phased one. In this case $\ell=0\rightarrow \ell=2$ conversion is achieved \cite{yao2011orbital}.}
	\label{fig:SPP}
\end{figure}

As future communication networks aim for unprecedented performance, the precise spatial mode-selective manipulation of OAM carriers—including detection, (de)multiplexing, sorting, and switching—becomes a critical enabler for high-speed optical systems. Among the various methods for generating and detecting OAM modes, spiral phase plates (SPPs) offer a convenient and efficient approach. As displayed in Figure \ref{fig:SPP}, when a Gaussian beam passes through an SPP, the resulting helical phase profile imparts an OAM mode with topological charge $\ell$. The thickness profile of the SPP is designed as $\ell \lambda \phi/2\pi(n-1)$, where $n$ is the refractive index of the medium
and $\lambda$ is the wavelength \cite{fatkhiev2021recent}.

SPPs boast several advantages, including high conversion efficiency, compatibility with high-power lasers and millimeter wavelengths, and ease of implementation without requiring specialized laser sources or propagation direction deviation. However, they also have some limitations: SPPs generate only single OAM modes, require precise fabrication, and lack versatility in detecting composite modes. Another notable point about SPPs is that, based on reversibility, OAM modes can be detected using an inverse-SPP designed for the corresponding modes. However, inverse-SPPs can detect only a specific OAM mode, necessitating the use of beam splitters and multiple inverse-SPPs for composite mode detection. Despite these challenges, SPPs remain a popular choice for single-mode applications due to their simplicity, affordability, high precision, high conversion efficiency, specific working frequency, and market availability, particularly when mode purity and processing complexity are not critical
\cite{oemrawsingh2004production,yao2011orbital,schemmel2014modular,willner2015optical,chen2019orbital}.

Motivated by these considerations, this paper proposes an OAM detection design leveraging inverse-SPP for free-space communication channels. In real-world scenarios, factors such as atmospheric turbulence and misalignments can induce lateral displacements and angular phase distortions, reducing the efficiency of communication links can induce lateral displacements and angular phase distortions, reducing the efficiency of the communication link \cite{willner2017recent}. While these impairments are absent in ideal systems, inherent inefficiencies in OAM conversion by SPPs persist. This study investigates these inefficiencies by analyzing the optical efficiency, crosstalk, and signal-to-interference ratio (SIR) parameters, through both analytical and numerical approaches. Specifically, Laguerre-Gaussian (LG) modes, a key class of OAM-carrying beams, are employed as a basis for the decomposition of beams in the transmitter, receiver, and during free-space propagation \cite{allen1992orbital}. The detector's design has been optimized in such a way to minimize undesired mode power (crosstalk) while maximizing the desired mode power (efficiency).

The rest of the paper is organized as follows: Section \ref{sec: Theory} introduces LG modes and their role in mode expansion, while the theoretical framework for OAM detection using inverse-SPPs is detailed in included subsections, with analytical investigations for two different scenarios. The OAM detection by SPP is also numerically investigated in Sections \ref{Sec: Num}. Section \ref{Sec: Conclusion} discusses the results, including efficiency and cross-talk performance, and summarizes the key findings and contributions. 
 
\section{Analytical investigation}
\label{sec: Theory}
Paraxial beams, such as Gaussian beams, are characterized by their slowly-varying envelope along the optical axis \cite{siegman1986lasers}. These beams provide a fundamental solution to the paraxial Helmholtz wave equation in cylindrical coordinates ($\rho, \phi, z$). The complex amplitude of Gaussian modes, which describe the transverse field distribution, is mathematically expressed as \cite{saleh2019fundamentals}:

\begin{equation}
U\left(\rho,z\right) = \sqrt{\frac{2}{\pi}} \frac{1}{W(z)} \exp{\left(\frac{-\rho^2}{W^2(z)}\right)}
\exp{\left[-jk\left(z+\frac{\rho^2}{2R(z)}\right)+j\zeta(z)\right]},
\label{eq:Gaussian_beam}
\end{equation}
where $W(z) = W_0\sqrt{1+(z/z_0)^2}$ represents the beam width, $R(z) = z[1+(z_0/z)^2]$ indicates the wavefront's radius of curvature, and $\zeta(z) = \tan^{-1}(z/z_0)$ is due to Gouy phase shift. Also, $W_0$ and $z_0$ are the beam's waist radius and Rayleigh range, respectively. This equation indicates that Gaussian modes lack any dependence on the azimuthal angle, $\phi$, signifying that they do not inherently carry OAM. However, it will be demonstrated that a Gaussian beam can acquire OAM after passing through an SPP.

Laguerre-Gaussian ($\text{LG}_{\ell,p}$) modes represent another set of solutions to the paraxial Helmholtz wave equation in cylindrical coordinates. These modes are characterized by two indices: the azimuthal index $\ell$, which corresponds to the orbital angular momentum of the mode, and the radial index $p$, which defines the radial structure of the intensity distribution. Their complex amplitude is expressed as \cite{andrews2012angular,saleh2019fundamentals}:
\begin{equation}
\text{LG}_{\ell,p}(\rho,\phi,z) = A_{\ell,p} \frac{1}{W(z)} \left(\frac{\rho\sqrt{2}}{W(z)}\right)^{|\ell|} L_p^{|\ell|}\left(\frac{2\rho^2}{W^2(z)}\right)
\exp{\left(\frac{-\rho^2}{W^2(z)}\right)}
\exp{\left[-jk\left(z+\frac{\rho^2}{2R(z)}\right)-j\ell\phi+j(|\ell|+2p+1)\zeta(z)\right]}.
\label{eq:LG_beam}
\end{equation}
Here, $L_p^{|\ell|}(.)$ represents the generalized Laguerre polynomial, and $A_{\ell,p} = \sqrt{2p!/\left(\pi \left(p+|\ell| \right)! \right)}$ is a normalization factor to ensure the power of the beam is equal to unity. These normalized definitions of Gaussian and LG beams (Eqs. \eqref{eq:Gaussian_beam} and \eqref{eq:LG_beam}) are applied in all the calculations of this study. Table \ref{tab:1} demonstrates the intensity distributions and phase patterns of several $\text{LG}_{\ell,p}$ modes. It depicts that the intensity distribution of different values of $p$ consists of $p+1$ concentric rings, while the number of $2\pi$ phase variations and the radius of the doughnut-shaped intensity distribution increase with enhancing $\ell$.

\begin{table}[h]
    \centering
    \caption{Intensity distributions and phase patterns of several $\text{LG}_{\ell,p}$ modes with different ${\ell}$ and $p$ orders.}
    \includegraphics[width=0.6\linewidth]{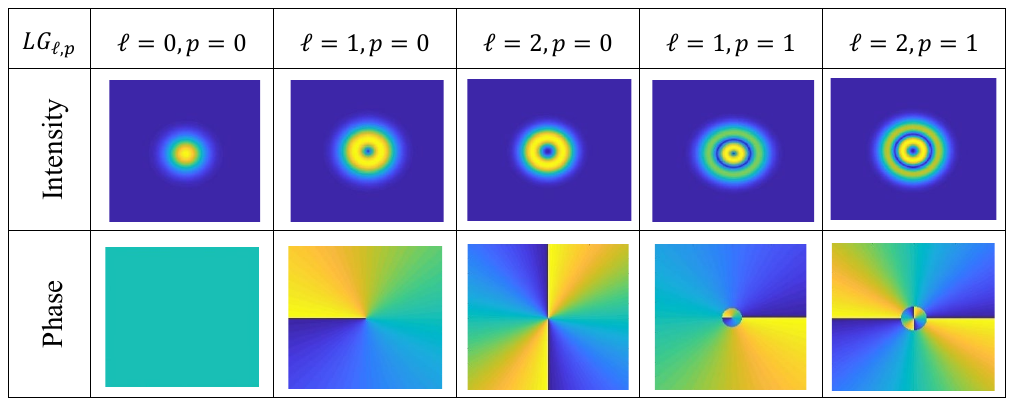}
    \label{tab:1}
\end{table}

Since $\text{LG}_{\ell,p}$ modes form a complete orthogonal basis, they are commonly used to decompose a general OAM-carrying beam, $U(\rho,\phi,z)$, as a weighted superposition of LG modes. Mathematically, this decomposition is expressed as:

\begin{equation}
U(\rho,\phi,z) = \sum\limits_{m,n} C_{m,n} f_{m,n}(\rho,\phi,z),
\label{eq:decomposition}
\end{equation}
where
\begin{equation}
f_{m,n}(\rho,\phi,z) = \frac{1}{W(z)}\sqrt{\frac{2n!}{\pi(n+|m|)!}} \Psi_n^m(\rho,z) \Phi_m(\phi)
\exp{\left[-jkz+j(|m|+2n+1)\zeta(z)\right]}, 
\label{eq:basis}
\end{equation}
and
\begin{gather}
\Psi_n^m(\rho,z) = \left(\frac{\rho\sqrt{2}}{W(z)}\right)^{|m|} L_n^{|m|}\left(\frac{2\rho^2}{W^2(z)}\right)
\exp{\left(\frac{-\rho^2}{W(z)^2}\right)} \exp{\left(-jk\frac{\rho^2}{2R(z)}\right)},\\
\Phi_m(\phi) = \exp{(-jm\phi)}.
\end{gather}
In these equations, the indices $m$ and $n$ are used as azimuthal ($\ell$) and radial ($p$) indicators, respectively. Using Eq. (\ref{eq:decomposition}), the generation and detection of OAM beams via SPP ‌are analytically examined in Sections \ref{Sec: gen} and \ref{Sec: detection}. 

\subsection{OAM beams generation using ideal SPP}
\label{Sec: gen}

In an SPP with a phase profile of $e^{-j\ell\phi}$, where $\ell$ is the topological charge, an incident Gaussian beam is transformed into a beam with a helical phase structure, which is a defining characteristic of an OAM-carrying beam. According to Eq. \eqref{eq:decomposition}, the generated OAM beam can be decomposed into LG modes. For a Gaussian beam with normalized power incident on the SPP, the beam is converted into a superposition of LG modes, where the corresponding amplitudes are determined by: 

\begin{equation}
\left|C_{\ell,n}\right| = \frac{|\ell|}{2\sqrt{n!(|\ell|+n)!}}\Gamma{\left(\frac{|\ell|}{2}+n\right)},
\label{eq:OAM_generation}
\end{equation}
where $\Gamma{(.)}$ denotes the Gamma function. These coefficients indicate that for an ideal SPP with no imperfections, the output beam’s azimuthal index is $m = \ell$, while all other azimuthal indices are zero. The detailed derivation of Eq. \eqref{eq:OAM_generation} is provided in Appendix \ref{App: App_A}. Figure \ref{fig:OAM_generation} illustrates the amplitudes of radial indices $(n)$ of the OAM beams generated by SPPs with $\ell = 1$, $\ell = 3$, and $\ell = 5$ topological charges, decomposed into LG modes. The figure highlights that the probability of generating lower-order modes (modes with smaller radial indices) through an SPP is higher than that of generating higher-order modes (modes with larger radial indices). Furthermore, as the topological charge $(\ell)$ of the SPP increases, the contribution of higher-order modes becomes more prominent.
\begin{figure}[t]
	\centering
	\includegraphics[width=0.5\linewidth]{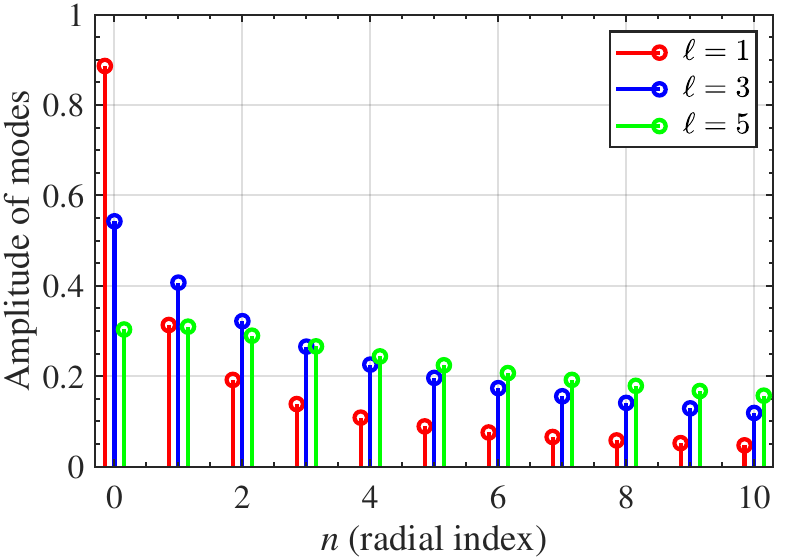}
	\caption{The amplitudes of radial indices ($n$) of decomposed generated OAM beams (Gaussian beam passed through an SPP) into $\text{LG}_{\ell,n}$ modes, via SPPs with $\ell = 1$, $\ell = 3$, and $\ell = 5$ topological charges.}
	\label{fig:OAM_generation}
\end{figure}

\subsection{OAM beams detection using ideal SPP}
\label{Sec: detection}
In a manner analogous to the beam generation, an SPP with an inverse topological charge can be employed to detect an OAM mode. However, as highlighted in the previous section, the generation of OAM beams by passing a Gaussian beam through an SPP does not yield a pure LG mode. For instance, when an SPP with a topological charge of $\ell_0$ is used to generate an OAM beam, several $\text{LG}_{\ell_0,n}$ modes with different radial indices ($n$) are produced. Therefore, in the detection process using an SPP, two distinct scenarios are investigated:

1. Detection of pure LG modes:

In this scenario, the transmitted modes are pure $\text{LG}$ modes. Two modes with distinct azimuthal indices $(\ell_1, \ell_2)$ but identical radial indices ($p$) are sent from the transmitter. After propagation through free space, these modes encounter an SPP with inverse topological charge $\left(\mathrm{SPP}_{-\ell_1}\right)$ at the receiver. Subsequently, the beams are passed through a circular aperture of radius $R$, positioned at a distance $d$ from the SPP.

2. Detection of OAM beams generated by Gaussian beams passed through an SPP:

In this scenario, two Gaussian beams pass through two different SPPs with topological charges $(\ell_1, \ell_2)$ at the transmitter. After free-space propagating, these OAM beams are detected at the receiver by an SPP with inverse topological charge ($\left(\mathrm{SPP}_{-\ell_1}\right)$), followed by passage through a circular aperture of radius $R$ located at a distance $d$ from the SPP.

\begin{figure}[h]
	\centering
	\includegraphics[width=0.5\linewidth]{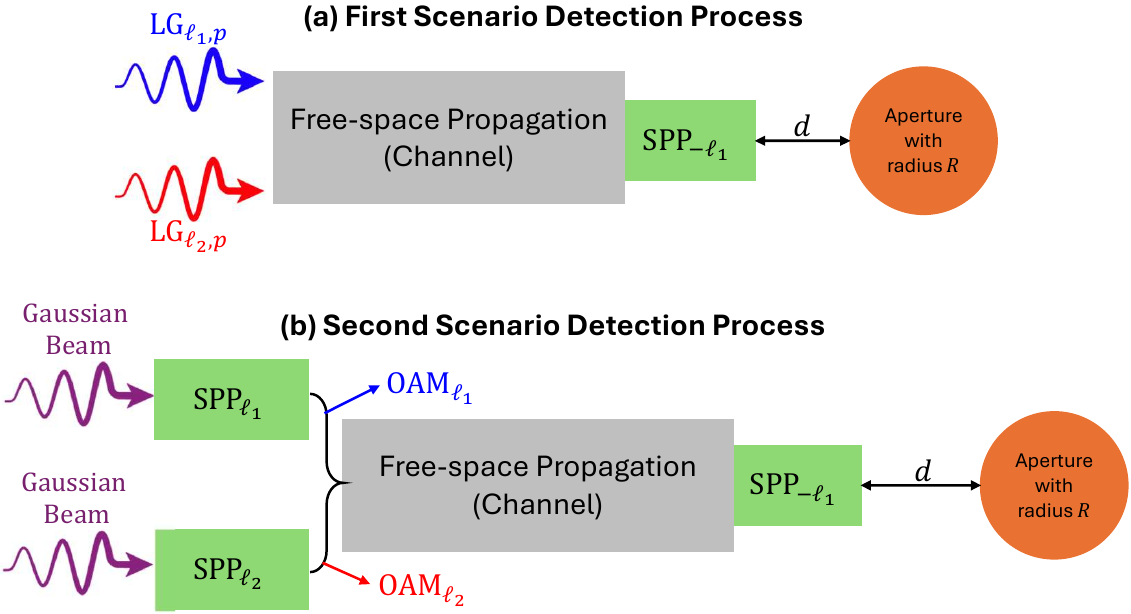}
	\caption{Illustration of the two proposed detection scenarios for OAM beams using an SPP with an inverse topological charge. (a) In the first scenario, it was assumed that pure LG modes were transmitted. (b) In the second scenario, OAM modes were generated by Gaussian beams passing through an SPP and then transmitted.}
	\label{fig:Scenarios}
\end{figure}

Figure \ref{fig:Scenarios} depicts these two proposed scenarios for OAM-carrying beams detection. The second scenario represents a more practical and realistic approach, as generating pure LG modes is significantly more complicated than generating Gaussian beams, which are typically the outputs of lasers.

In both scenarios, the transmitted modes inevitably mix during free-space propagation. The key objective at the receiver is to separate these modes based on the orthogonality of OAM modes. Notably, the same channel—operating at the same time and frequency—is utilized for propagating both modes from transmitter to receiver, without employing additional multiplexing techniques such as FDM or TDM.

\subsubsection{First Scenario: Detection of pure LG modes}
In this scenario, for simplicity, it is assumed that the received beams at the detector are pure LG modes. Accordingly, for the received LG mode with azimuthal index $\ell$ passing through an SPP with topological charge $-\ell$, the resulting beam loses its helical phase, becoming a non-rotating beam. Thus, if the received beam is generally an LG mode with a radial index of $p_i$ and azimuthal index of $\ell_i$ (i.e., $\text{LG}_{\ell_i,p_i}$) and SPP has the topological charge of $-\ell_0$, then the output beam after passing through the SPP can be expressed as a superposition of LG modes. Using the explicit form of the Laguerre function, the amplitude of each resulting LG mode is written as the equation (see the derivation in Appendix \ref{App: App_B}):

\begin{equation}
\begin{split}
\left|C_{\ell_i-\ell_0,n}\right| = \frac{\sqrt{p_i!(p_i + |\ell_i|)!}}{\sqrt{(|\ell_i-\ell_0|+n)!n!}} 
\sum\limits_{m=0}^{p_i} \left[\frac{(-1)^m}{(p_i-m)!(|\ell_i|+m)!m!} \left(\frac{|\ell_i-\ell_0|}{2} - \frac{|\ell_i|}{2}-m\right)_n\Gamma\left(\frac{|\ell_i-\ell_0|}{2} + \frac{|\ell_i|}{2}+m+1\right)\right].
\label{eq:LG_amplitudes}
\end{split}
\end{equation}
Here, $(.)_n$, known as the Pochhammer symbol or the rising factorial, is defined as $(x)_n = \Gamma(x+n)/\Gamma(x)$, where $\Gamma(x)$ represents the Gamma function. From the equation, it can be observed that if $\ell_i=\ell_0$, the output beam will have no helical phase, as expected. However, to enhance detection performance, an optical element should be utilized to filter out undesired modes following the SPP. The design of an optimal detector focuses on maximizing detection efficiency while minimizing crosstalk. Here, efficiency is defined as the power of desired modes, which are the LG modes with zero azimuthal index ($\ell=0$) after passing through the SPP. Crosstalk refers to the power of undesired modes, which are the other LG modes carrying nonzero azimuthal indices ($\ell \neq 0$). Due to the normalization factors defined in Eqs. \eqref{eq:Gaussian_beam} and \eqref{eq:LG_beam}, the power of the incident waves is set to one. As a result, the efficiency directly corresponds to the power of the desired modes without requiring further normalization. Similarly, crosstalk is directly calculated as the power of the undesired modes. Accordingly, the total intensity of desired modes after detection is obtained as
\begin{equation}
I_{\text{eff}}(\rho,z) = \left|\sum\limits_{n} C_{0,n} \text{LG}_{0,n}(\rho,z)\right|^2.
\label{eq:I_eff}
\end{equation}

A circular aperture is considered for detection following the SPP in accordance with practical detectors. The aperture radius and propagation distance between the SPP and aperture $(d)$ should be optimized for more optimal detection. Consequently, the power associated with the desired modes is defined as:
\begin{equation}
P_{\text{eff}} = \int_A I_{\text{eff}}(\rho,z)p(\rho)dA,
\label{eq:P_eff_def}
\end{equation}
where $p(\rho)$ is the transfer function of a circular aperture with radius $R$ and is defined as:
\begin{equation}
p(\rho) = \left\{
\begin{array}{ll}
1 & \text{inside the aperture}\:(\rho \leq R)\\
0 &	\text{outside the aperture}\:(\rho>R).
\end{array} \right.
\end{equation}

Using the explicit form of the Laguerre function in Eq. \eqref{eq:I_eff}, the power of desired modes is calculated as (see the derivation in Appendix \ref{App: App_C}),
\begin{equation}
P_{\text{eff}} \approx \frac{\pi}{2} \sum\limits_{n} \left|C_{0,n}\right|^2 A_{0,n}^2 \sum\limits_{m=0}^n a_m \sum\limits_{k=0}^n a_k \gamma\left({m+k+1,\frac{2R^2}{W^2(z)}}\right),
\label{eq:P_eff1}
\end{equation}
where $\gamma(.)$ denotes the incomplete gamma function and $a_i$ is defined as $a_i = n!/\left((n-i)!(i!)^2\right)$. Additionally, $R$ is the radius of the circular aperture used at the detector, and $W(z)$ shows the beam width which determines the optimum propagation distance of the beams between SPP and the aperture. As previously described, crosstalk refers to the power of remaining LG modes with nonzero azimuthal indices. Therefore, in a similar manner, the crosstalk can be evaluated as
\begin{equation}
P_{\text{XT}} \approx \frac{\pi}{2} \sum\limits_{n} |C_{\ell_i-\ell_0,n}|^2 A_{\ell_i-\ell_0,n}^2 \sum\limits_{m=0}^n b_m \sum\limits_{k=0}^n b_k \gamma\left({m+k+|\ell_i-\ell_0|+1,\frac{2R^2}{W^2(z)}}\right).
\label{eq:P_XT1}
\end{equation}

In this equation, $b_j$ is defined as $b_j = (n+|\ell_i-\ell_0|)!/\left((n-j)!(j +|\ell_i-\ell_0|)! j!\right)$. 

\begin{figure}[h]
	\centering 
	\includegraphics[width=0.5\linewidth]{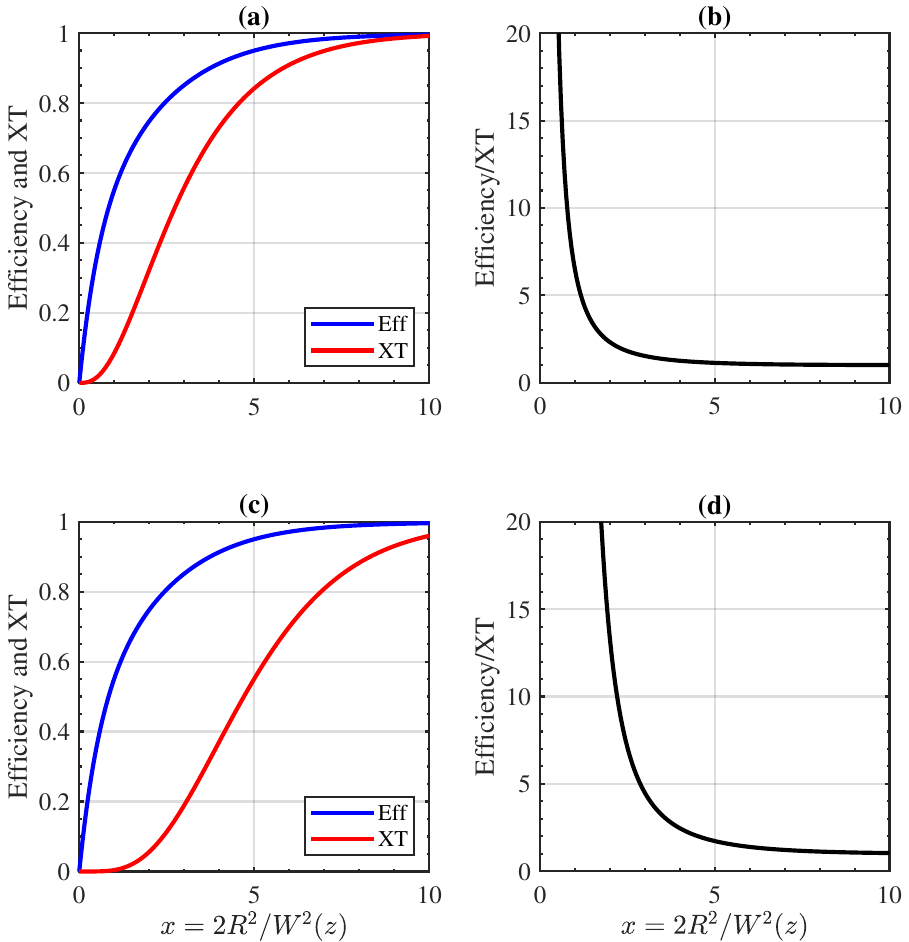}
	\caption{The efficiency and crosstalk (left) and their ratio (right) versus $x = 2R^2/W^2(z)$ parameter for detection of (a),(b) $\text{LG}_{\ell_1 = 1 ,p = 0}$, $\text{LG}_{\ell_2 = 3 ,p = 0}$ and (c),(d) $\text{LG}_{\ell_1 = 1 ,p = 0}$, $\text{LG}_{\ell_2 = 5 ,p = 0}$ incident modes, using an SPP with $\ell = -1$.}
	\label{fig:LG_detection1}
\end{figure}

Figure \ref{fig:LG_detection1} illustrates the efficiency and crosstalk (left) and their ratio (right), as the signal-to-interference ratio (SIR) parameter, with respect to $x = 2R^2/W^2(z)$ parameter (which is a function of the circular aperture radius and the propagation distance between SPP and aperture) for different azimuthal indices. In Fig. \ref{fig:LG_detection1} (a) and (b), $\text{LG}_{\ell_1 = 1 ,p = 0}$ and $\text{LG}_{\ell_2 = 3 ,p = 0}$ modes are assumed as the incident ones while SPP with topological charge $\ell = -1$ is used for their detection. Fig. \ref{fig:LG_detection1} (c) and (d) also considers $\text{LG}_{\ell_1 = 1 ,p = 0}$ and $\text{LG}_{\ell_2 = 5 ,p = 0}$ incident modes with the same SPP. The comparison of the left plots highlights that while the difference in azimuthal indices of the desired and undesired modes increases, the crosstalk (XT) is reduced. This implies that modes with a greater azimuthal index difference are easier to distinguish and detect effectively. The right plots also point out that the best detection corresponding to the highest SIR is achieved when the circular aperture radius tends to zero, which is not physically feasible. However, the comparison indicates that when the difference in azimuthal indices is increased, the same SIR is achieved at a higher $x$ parameter, which means a larger aperture radius would be needed, making it more practical for detection.

\begin{figure}[t]
	\centering
	\includegraphics[width=0.5\linewidth]{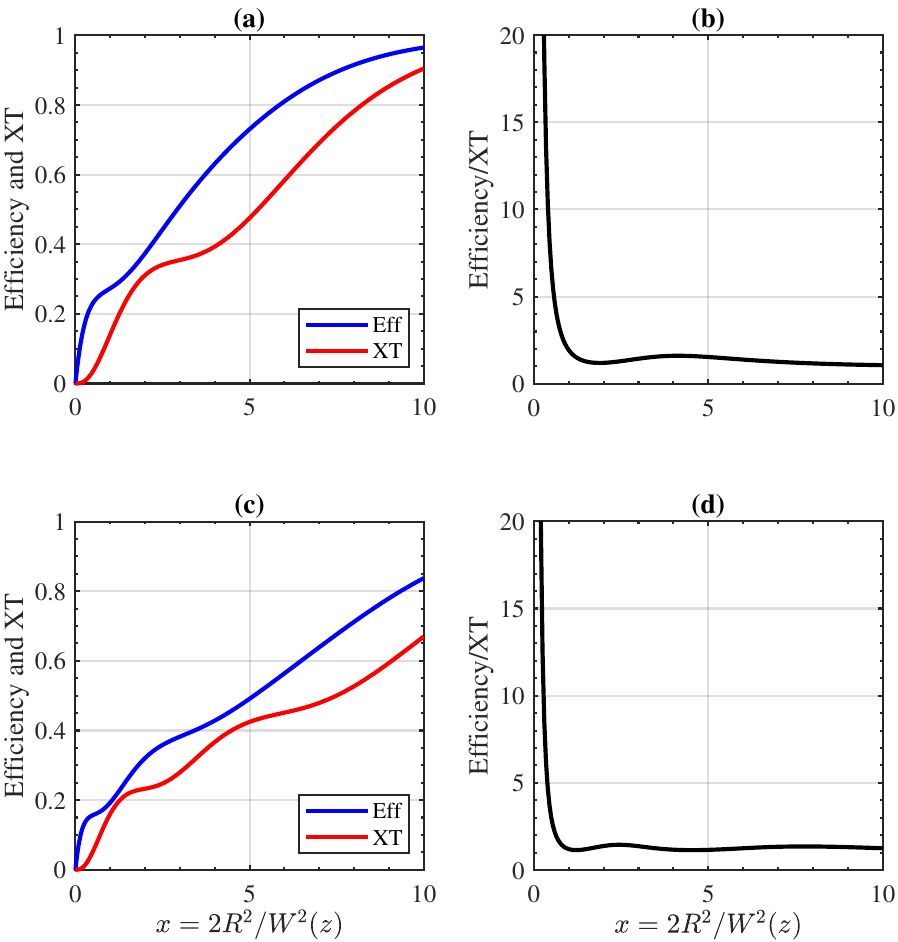}
	\caption{The efficiency and crosstalk (left) and their ratio (right) versus $x = 2R^2/W^2(z)$ parameter for detection of (a),(b) $\text{LG}_{\ell_1 = 1 ,p = 1}$, $\text{LG}_{\ell_2 = 3 ,p = 1}$  and (c),(d) $\text{LG}_{\ell_1 = 1 ,p = 2}$ and $\text{LG}_{\ell_2 = 3 ,p = 2}$ incident modes, using an SPP with $\ell = -1$ topological charge.}
	\label{fig:LG_detection2}
\end{figure}

Moreover, to investigate the effect of radial indices of incident modes on their detection by SPP, Fig. \ref{fig:LG_detection2} represents the variations of efficiency and crosstalk (left) and their ratio (right) versus $x$ parameter; the top plots correspond to $\text{LG}_{\ell_1 = 1 ,p = 1}$ and $\text{LG}_{\ell_2 = 3 ,p = 1}$, while the bottom plots correspond to $\text{LG}_{\ell_1 = 1 ,p = 2}$ and $\text{LG}_{\ell_2 = 3 ,p = 2}$ incident modes, while both detections are considered with the same SPP of topological charge $\ell = -1$.
The comparison of Fig. \ref{fig:LG_detection1} and Fig. \ref{fig:LG_detection2} implies that when the radial index of the incident LG beam is increased, both efficiency and crosstalk are generally decreased and their overall enhancements by $x$ parameter, are associated with undulations complied with the numbers of radial indices (and hence to the number of rings in power distributions of Table \ref{tab:1}). On the other hand, the corresponding size of the circular aperture for achieving better detection (higher SIR) is decreased for higher radial indices. However, tiny peaks appear in Fig. \ref{fig:LG_detection2}(b) and (d), which are associated with the nonzero radial indices used. These peaks represent local maxima of efficiency over crosstalk as the $x$ parameter varies and could be considered optimal points for detection. Nonetheless, since generating pure LG modes with nonzero radial indices is challenging, this approach may not be a promising practical solution for designing an optimal detector.
 
\subsubsection{Second Scenario: Detection of OAM Beam}
In this second scenario, a more practical and realistic detection setup is considered. Unlike the first scenario, where pure LG modes were assumed to be transmitted and received, this approach involves OAM-carrying beams generated by passing Gaussian beams through an SPP. These beams then propagate through free space from the transmitter to the receiver. By accounting for the complexities and imperfections inherent in realistic beam generation and propagation, this scenario provides a comprehensive evaluation of the transceiver system's performance. 

\begin{figure}[h]
	\centering
	\includegraphics[width=0.5\linewidth]{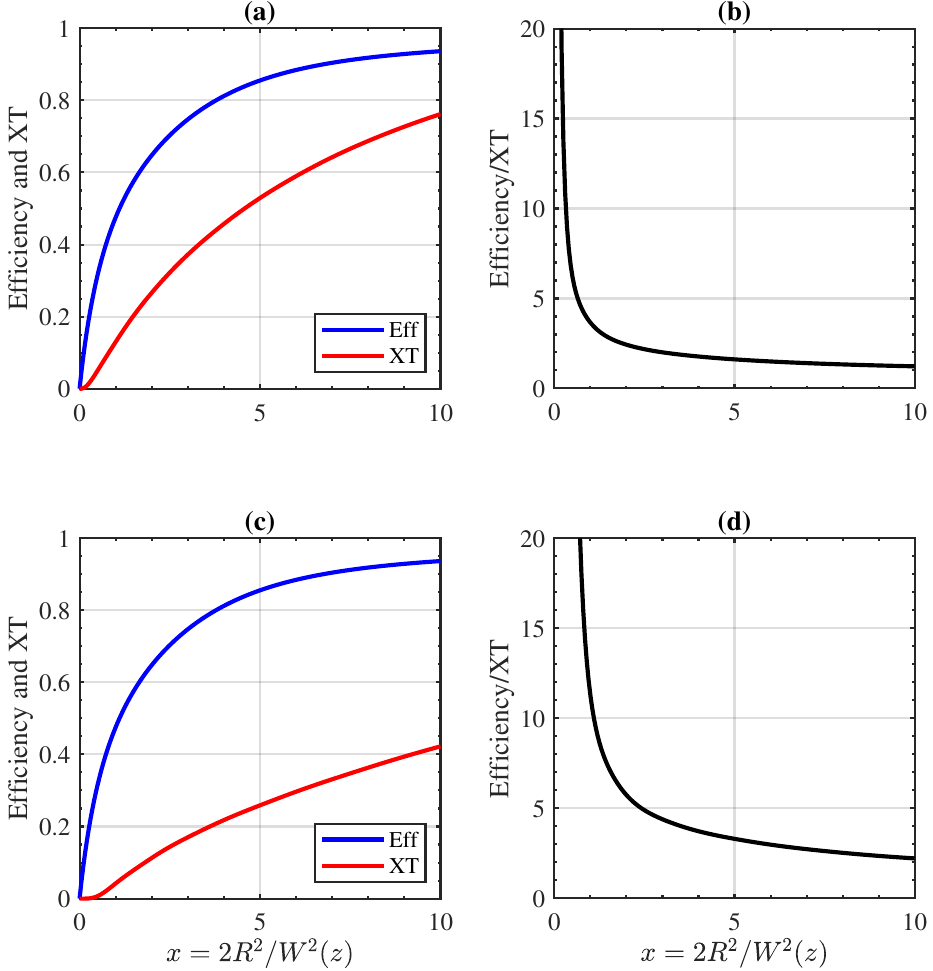}
	\caption{The efficiency and crosstalk (left) and their ratio (right) versus $x = 2R^2/W^2(z)$ parameter for detection of (a),(b) $\text{OAM}_{\ell_1 = 1}$, $\text{OAM}_{\ell_2 = 3}$ and (c),(d) $\text{OAM}_{\ell_1 = 1}$, $\text{OAM}_{\ell_2 =5 }$ incident modes, using an SPP with $\ell = -1$ topological charge.}
	\label{fig:OAM_detection}
\end{figure}

According to Section \ref{Sec: gen}, the OAM beam generated by an SPP is not a single LG mode. Instead, they can be expressed on the basis of LG modes. Therefore, if two OAM beams are generated using two SPPs with topological charges $\ell_1 $ and $\ell_2$, their respective decompositions can be expressed as
\begin{equation}
\text{OAM}_{\ell_1}(\rho,\phi,z) = \sum\limits_{n} C_{\ell_1,n}\text{LG}_{\ell_1,n}(\rho,\phi,z),
\quad
\text{OAM}_{\ell_2}(\rho,\phi,z) = \sum\limits_{n} D_{\ell_2,n}\text{LG}_{\ell_2,n}(\rho,\phi,z),
\end{equation} 
where $C_{\ell_1,n}$ and $D_{\ell_2,n}$ are defined in Eq. \eqref{eq:OAM_generation}. The generated OAM beams are transmitted to the receiver, where they first pass through an SPP with the topological charge of $-\ell_1$ and then reach a circular aperture with radius $R$ for detection. In a similar manner to the first scenario, the efficiency and crosstalk are calculated as (see the derivation in Appendix \ref{App: App_D}),
\begin{equation}
\begin{split}
P_{\text{eff}} \approx& \sum\limits_{p} \sum\limits_{n} \left[\frac{|\ell_1|}{2n!(|\ell_1| + n)!} \Gamma\left(\frac{|\ell_1|}{2} + n\right)\sum\limits_{m = 0}^{p} \frac{(-1)^m p!}{(p-m)!(m!)^2} \left({\frac{|\ell_1|}{2} - m}\right)_n \Gamma\left(\frac{|\ell_1|}{2} + m + 1\right)\right]^2 \\ 
&\sum\limits_{\ell = 0}^{p} a_l \sum\limits_{k = 0}^{p} a_k \gamma\left(\ell+k+1,\frac{2R^2}{W^2(z)}\right),
\end{split}
\label{eq:P_eff2}
\end{equation}
and
\begin{equation}
\begin{split}
P_{\text{XT}} &\approx \sum\limits_{p} \left[\frac{p!}{(|\ell_2-\ell_1| + p)!}\right]^2 \sum\limits_{\ell = 0}^{p} b_\ell \sum\limits_{k = 0}^{p} b_k \gamma\left(\ell+k+|\ell_2 - \ell_1| + 1,\frac{2R^2}{W^2(z)}\right) \\ &\sum\limits_{n} \left[\frac{|\ell_2|}{2n!(|\ell_2| + n)!}  \Gamma\left(\frac{|\ell_2|}{2} + n\right)\sum\limits_{m = 0}^{p} \frac{(-1)^m (p + |\ell_2 - \ell_1|)!}{(p-m)!(|\ell_2 - \ell_1| + m)!m!}
\left({\frac{|\ell_2 - \ell_1|}{2} -\frac{|\ell_2|}{2}- m}\right)_n \Gamma\left(\frac{\left|\ell_2 - \ell_1\right|}{2} +\frac{|\ell_2|}{2} + m + 1\right)\right]^2,
\end{split}
\label{eq:P_XT2}
\end{equation}
where $n$ and $p$ indicate the radial indices resulting from the decomposition at the transmitter and receiver, respectively. Similar to the previous scenario, the efficiency, crosstalk, and their ratio are plotted against the $x$ parameter for different azimuthal indices. Figure \ref{fig:OAM_detection} illustrates the mentioned parameters, the top plots correspond to $\text{OAM}_{\ell_1 = 1}$ and $\text{OAM}_{\ell_2 = 3}$ incident beams, when an SPP with topological charge of $\ell = -1$ is used, and the bottom plots correspond to $\text{OAM}_{\ell_1 = 1}$ and $\text{OAM}_{\ell_2 = 5}$ incident beams using the same SPP. These plots reveal that although similar to the previous scenario both the efficiency and crosstalk increase with the $x$ parameter, but in this case, which considers the general OAM-carrying beams (not only $\text{LG}$ modes), their enhancements are intuitively smoother. In other words, the impact of radial indices is less pronounced, especially in comparison to Fig. \ref{fig:LG_detection2}. Analogously, as the difference between the azimuthal indices of the desired and undesired modes increases, the crosstalk decreases, enhancing the detection quality. Also, the aperture size required to achieve equal SIR increases, making the detection more practical for larger azimuthal differences.

The results of this scenario closely resemble those of the previous one, indicating that either scenario can be employed depending on the required accuracy. Both scenarios validate that a well-designed SPP can function effectively as a detector, efficiently isolating OAMs with distinct topological charges.

\subsection{Noise Considerations and Design Procedure}
Efficiency and crosstalk, while critical, cannot be regarded as the sole limiting factors in OAM detection; noise also plays a significant role. As demonstrated earlier, minimizing both efficiency and crosstalk and maximizing their ratio are achievable by reducing the aperture radius to near zero. However, this approach makes noise the dominant limiting factor in this type of detection.

To address this, a design procedure is proposed that incorporates noise considerations. Signal-to-interference-plus-noise ratio (SINR) and noise-equivalent power (NEP) are essential parameters to be defined at the receiver. Given the orthogonality of LG modes, it is reasonable to treat crosstalk analogously to noise. Accordingly, based on the SINR definition, a linear relationship between efficiency and crosstalk can be expressed as:
\begin{equation}
P_{\text{eff}} = \text{SINR}\left(P_{\text{XT}} + P_{\text{n}}\right),
\label{eq:SINR}
\end{equation}
where $P_{\text{n}}$ represents the noise power. According to Section \ref{Sec: detection}, efficiency and crosstalk can be plotted as functions of $x$. Subsequently, efficiency can be plotted against crosstalk. The intersection point of Eq. \eqref{eq:SINR} and the efficiency-crosstalk relationship determines the optimal efficiency and crosstalk values for each detection scenario. Consequently, the optimal value of $x$ can be found.

\section{Numerical investigation}
\label{Sec: Num}
This section presents a numerical analysis of the first and second scenarios to verify the findings of Section \ref{sec: Theory}. To begin, the propagation of a monochromatic optical wave with wavelength $\lambda$ in free space is examined. Considering the complex amplitude $U(x,y,z)$ between the input plane $(z=0)$ and output plane $(z=d)$, the free-space propagation maps $U(x,y,0)$ to $U(x,y,d)$ as a shift-invariant linear system. Such a system is characterized by its transfer function in the spatial frequency domain as $H(\nu_x, \nu_y) = \exp \left(-j2\pi d\sqrt{\lambda^{-2}-\nu_x^2-\nu_y^2} \right)$, where $\nu_x$ and $\nu_y$ represent the $x$ and $y$ directions, respectively \cite{saleh2019fundamentals, guo2020squeeze}. The free-space propagation process involves transforming the input plane to the spatial frequency domain $(\nu_x,\nu_y)$, multiplying it by the transfer function $H(\nu_x, \nu_y)$, and then converting the results back to the spatial domain $(x,y)$ \cite{goodman2005introduction}. Additionally, all optical elements including SPP are modeled by multiplying the input plane by the element transfer function.

Here, similar to Section \ref{sec: Theory}, the scenarios illustrated in Fig. \ref{fig:Scenarios} are numerically implemented. Following the first scenario described in Section \ref{Sec: detection}, two ideal LG modes (herein, $\text{LG}_{\ell_1=1,p=0}$ and $\text{LG}_{\ell_2=3,p=0}$) are considered as the input ones, to be detected by an SPP (here, $\text{SPP}_{\ell=-1}$). After propagation through a path of length $d$, the outputs are received at a circular aperture with radius $R$. Integrating the power over the aperture's surface obtains the designed detector's output power. Based on the charge identity of SPP ($\ell=-1$), the output powers corresponding to the input modes $\text{LG}_{\ell=1,p=0}$ and $\text{LG}_{\ell=3,p=0}$ determine the efficiency and crosstalk, respectively. The simulation parameters are $\lambda = 0.5\:\mu \mathrm{m}$ and $z_0 = 10\:\mathrm{m}$. The propagation distance between the transmitter and receiver is assumed to be $20\:\mathrm{m}$. For larger transmitter-to-receiver distances, paraxial beams would experience greater divergence in the transverse plane as they propagate along the $z$-axis, necessitating a larger transverse simulation plane. To simplify the simulations, however, the study focuses on this moderate propagation distance.

\begin{figure}[h]
	\centering
	\includegraphics[width=0.5\linewidth]{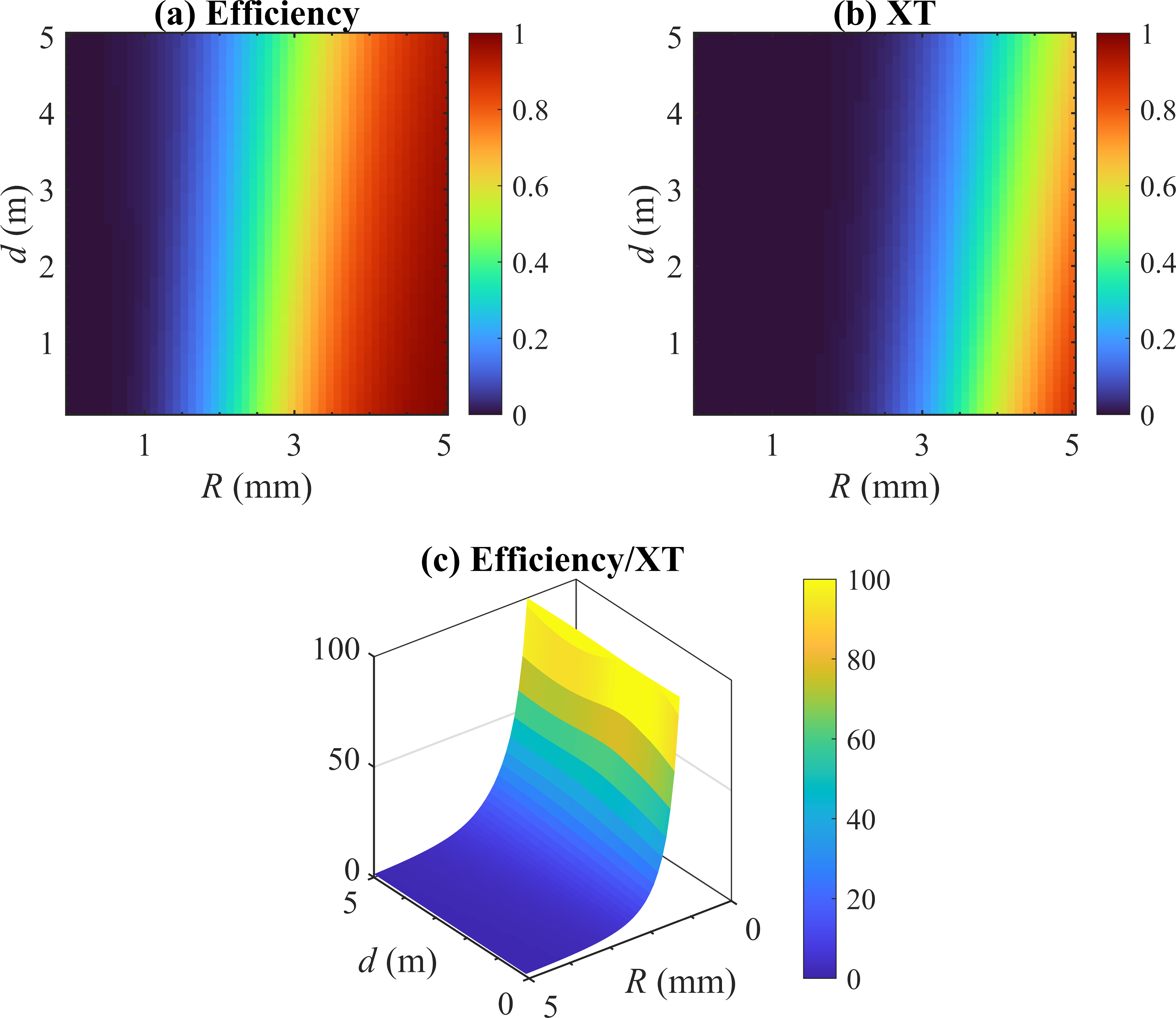}
        \caption{The efficiency (a), crosstalk (b), and their ratio (c) versus the radius of circular aperture, $R$, and the distance between SPP and aperture, $d$, for detection of $\text{LG}_{\ell_1 = 1 ,p = 0}$, $\text{LG}_{\ell_2 = 3 ,p = 0}$ incident modes, using an SPP with $\ell = -1$ topological charge.}	
        \label{fig:LG_detection_sim}
\end{figure}

Figure \ref{fig:LG_detection_sim}(a) and (b) respectively display the calculated efficiency and crosstalk, as a function of propagation distance between the SPP and the aperture ($d$) and aperture's radius ($R$). The results reveal that both efficiency and crosstalk increase with an increasing aperture radius and decrease with an increasing propagation distance, consistent with the findings in Fig. \ref{fig:LG_detection1}. Moreover, as shown in Fig. \ref{fig:LG_detection_sim}(c), the efficiency-to-crosstalk ratio (SIR) is maximized when the aperture radius tends to zero. Notably, the SIR does not exhibit significant dependence on the propagation distance ($d$).

\begin{figure}[h]
	\centering
	\includegraphics[width=0.5\linewidth]{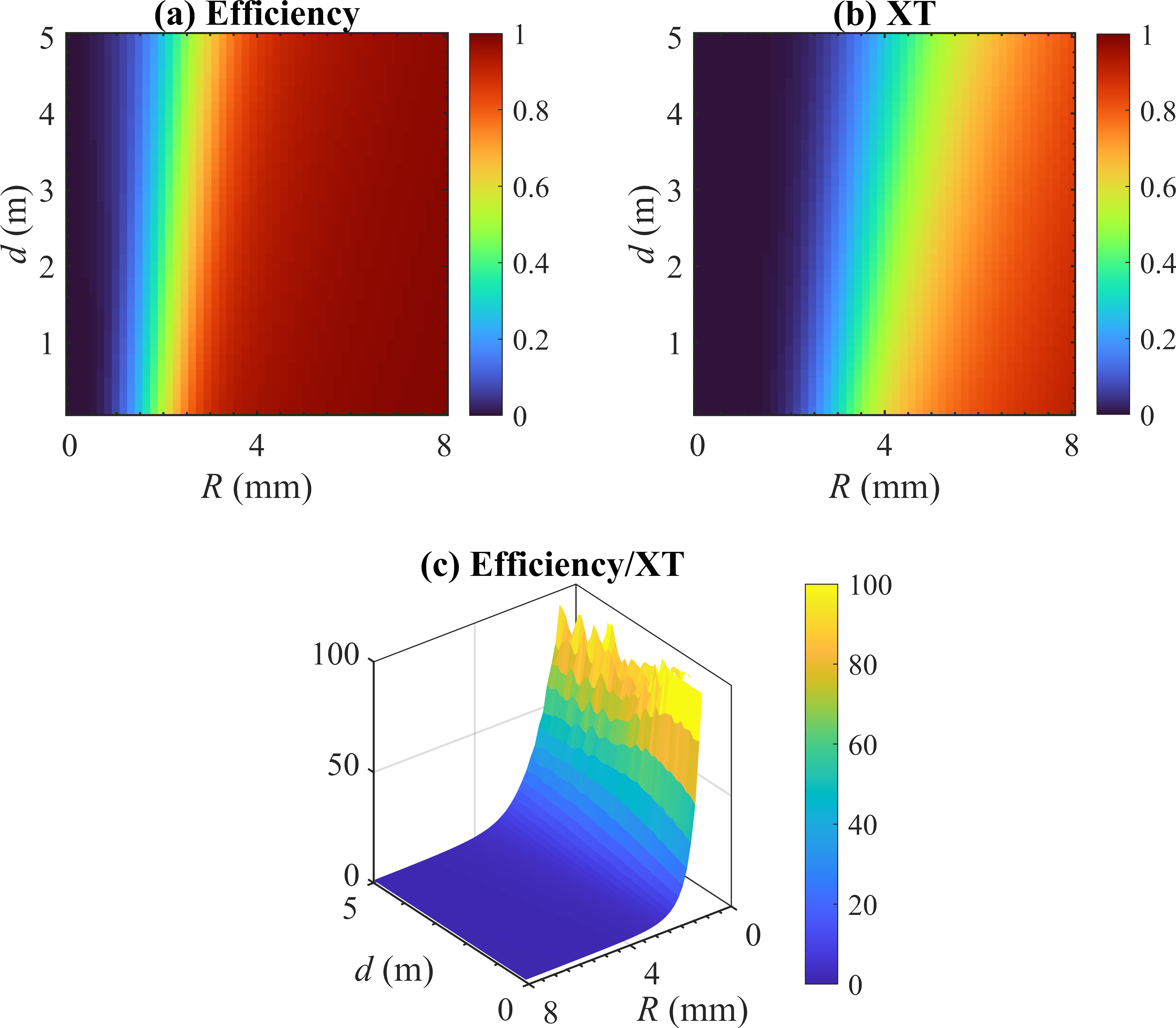}
        \caption{The efficiency (a), crosstalk (b), and their ratio (c) versus the radius of circular aperture, $R$, and the distance between SPP and aperture, $d$, for detection of $\text{OAM}_{\ell_1 = 1}$, $\text{OAM}_{\ell_2 = 3}$ incident modes, using an SPP with $\ell = -1$ topological charge.}	
        \label{fig:OAM_detection_sim}
\end{figure}

Similarly to the second scenario, two general OAM-carrying beams, $\text{OAM}_{\ell_1=1}$ and $\text{OAM}_{\ell_2=3}$, generated by an SPP, are considered as input modes in the scheme of Fig. \ref{fig:Scenarios}(b) with $\text{SPP}_{\ell=-1}$. The corresponding efficiency, crosstalk, and SIR are calculated and respectively presented in Fig. \ref{fig:OAM_detection_sim}(a-c) as functions of the aperture radius and propagation distance. Like the results for LG modes, the SIR is mainly unaffected by propagation length but increases at smaller aperture radii. Notably, higher SIR values are observed at larger aperture radii compared to the previous case, indicating that larger aperture sizes would be needed for detection in this scenario, making it a more practical approach.

The propagation distance between the SPP and the aperture at the receiver ranges from 0 to 5 meters, which is relatively large. To reduce this distance in practical detector implementations, a lens can be employed. Alternatively, the free space propagation can be substituted with a nonlocal optical device, which achieves the same functionality as free space while significantly reducing the physical length \cite{guo2020squeeze}.

\begin{figure}[h]
	\centering
	\includegraphics[width=0.5\linewidth]{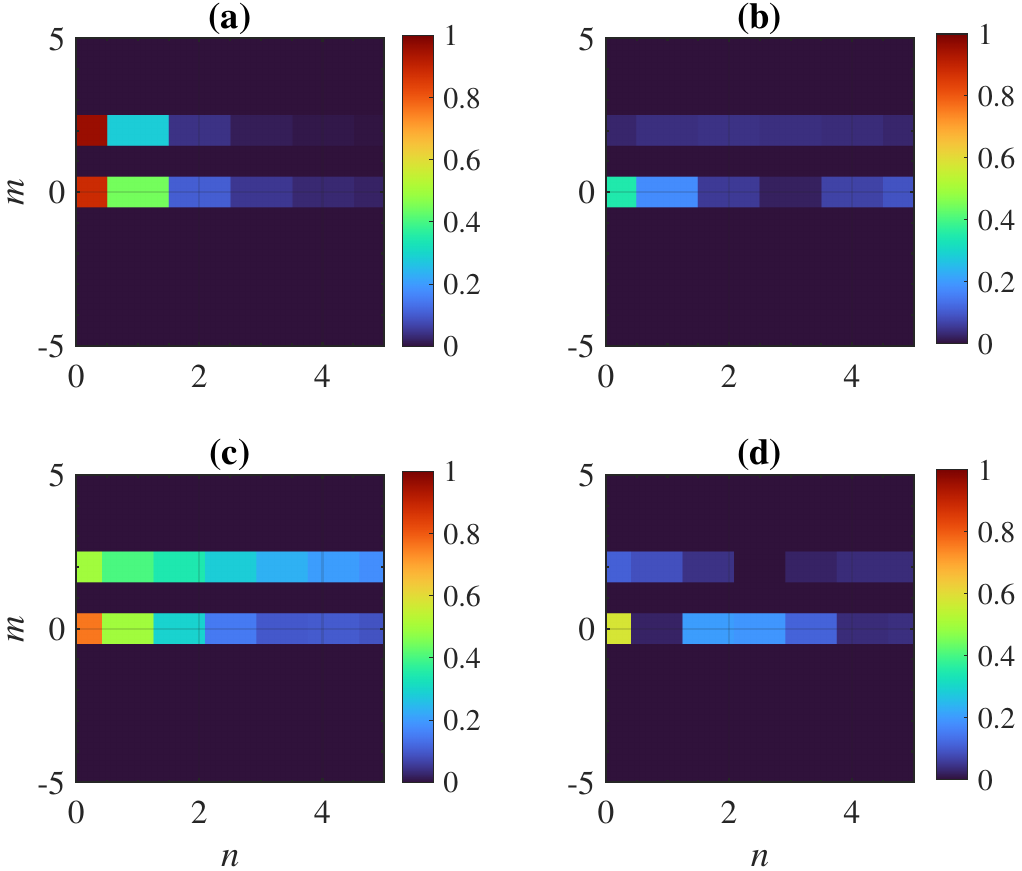}
	\caption{Top: the decomposition of the received LG beams (first scenario), passed through an SPP with inverse topological charge, based on LG modes: (a) before passing through the aperture and (b) after passing through the aperture for detection of $\text{LG}_{\ell_1 = 1 ,p = 0}$, $\text{LG}_{\ell_2 = 3 ,p = 0}$ incident modes, using an SPP with $\ell = -1$ topological charge. Bottom: the decomposition of the received OAM beams (second scenario), passed through an SPP with inverse topological charge, based on LG modes: (c) before passing through the aperture and (d) after passing through the aperture for detection of $\text{OAM}_{\ell_1 = 1}$, $\text{OAM}_{\ell_2 = 3}$ incident modes, using an SPP with $\ell = -1$ topological charge. Here, $m$ and $n$ denote the azimuthal and radial indices, respectively.}
	\label{fig:detection_decomposition}
\end{figure}

To evaluate the performance of the designed detector and the role of the aperture in filtering out undesired modes, the decomposition of received beams at the receiver, after passing through the SPP with inverse topological charge, is illustrated in Fig. \ref{fig:detection_decomposition} (where $m$ and $n$ denote the azimuthal and radial indices in the decomposition, respectively). Figure \ref{fig:detection_decomposition}(a) presents the LG modes decomposition (first scenario) received at the detector after passing through the SPP with inverse topological charge. As expected, the azimuthal indices $(m)$ of both beams are reduced by 1. Figure \ref{fig:detection_decomposition}(b) depicts the LG modes after passing through the aperture, highlighting a significant reduction in the amplitudes of undesired modes $(m=2)$ compared to their amplitudes before filtering. It is worth noting that the amplitudes of the desired modes $(m=0)$ also decrease slightly, as the aperture is not a perfect filter and cannot exclusively attenuate undesired modes. Nevertheless, the simple aperture used in this study demonstrates reasonable performance. Figures \ref{fig:detection_decomposition}(c) and (d) illustrate a similar filtering process for detection in the second scenario.

\section{Conclusion}
\label{Sec: Conclusion}
In this paper, a scheme for optimal detection of OAM-carrying modes is proposed which is based on SPPs with inverse topological charge to cancel the helical phase of the incident beam and an aperture for collection of desired mode power and filtering the undesired modes. Performance was evaluated by analyzing two incident modes: one with the identical inverse and the other with a different topological charge. Efficiency and crosstalk were quantified based on the optical power collected at the aperture, leading to the derivation of the SIR. The optimal detection condition was then determined by variation of SIR with respect to the propagation length and aperture radius. 

This approach leverages the low cost, high conversion efficiency, and tunability of SPPs. Additionally, integrating a lens can further enhance the compactness of the setup, enabling precise control over the propagation distance between the SPP and the receiver's aperture. While crosstalk remains a challenge, its impact is mitigated when detecting pure LG modes, particularly those with higher radial indices. Overall, the proposed method offers a straightforward, efficient, and reliable solution, paving the way for its implementation in future communication networks employing OAM modulation.

\appendix
\section{Derivation of Eq. (\ref{eq:OAM_generation})}
\label{App: App_A}
In this section, the generated OAM beam is decomposed into LG modes to determine the corresponding amplitudes. A Gaussian mode, after passing through an SPP, is described as
\begin{equation}
\text{OAM}_{\ell}(\rho,\phi,z) = \sqrt{\frac{2}{\pi}} \frac{1}{W(z)} \exp{\left(\frac{-\rho^2}{W^2(z)}\right)}
\exp{\left[-jk\left(z+\frac{\rho^2}{2R(z)}\right)+j\zeta(z) - j \ell \phi \right]}.
\label{eq:A1}
\end{equation}
The orthogonality of the defined basis functions in Eq. \eqref{eq:basis} (LG modes) arises from the orthogonality of the generalized Laguerre functions and azimuthal harmonic functions. Consequently, by defining $\psi = 2\rho^2/W^2(z)$, each coefficient in Eq. \eqref{eq:decomposition} for the decomposition of Eq. \eqref{eq:A1} can be computed using the following integral
\begin{equation}
C_{m,n} = \sqrt{\frac{n!}{(|m|+n)!}}\frac{1}{2\pi}\int_0^\infty \Psi_n^m(\psi)^*\exp\left(\frac{\psi}{2}\right) \exp(jm\phi)
\exp{\left[-jk\frac{\psi W^2(z)}{4R(z)}-j(|m|+2n)\zeta(z)-j\ell\phi\right]} d\phi d\psi.
\end{equation}
After some simplification, the following equation will be obtained.
\begin{equation}
C_{\ell,n} = \sqrt{\frac{n!}{(|\ell|+n)!}} \exp\left[-j(|\ell|+2n)\zeta(z)\right] \int_0^\infty \exp(-\psi) \psi^{\frac{|\ell|}{2}} L_n^{|\ell|}(\psi) d\psi.
\label{eq:A3}
\end{equation}
The integral in the above equation has a closed-form solution, which can be evaluated using the results provided in \cite{prudnikov1986integrals}, given by
\begin{equation}
\int_0^\infty \psi^{\alpha-1} \exp(-c\psi) L_n^\lambda(c\psi) d\psi= \frac{{(1-\alpha+\lambda)}_n}{n!c^\alpha}\Gamma(\alpha)
\label{eq:A4}.
\end{equation}
Using the above integral formula, $C_{\ell,n}$ in Eq. \eqref{eq:A3} is simplified as
\begin{equation}
C_{\ell,n} = \sqrt{\frac{n!}{(|\ell|+n)!}} \exp\left[-j(|\ell|+2n)\zeta(z)\right] \frac{(|\ell|/2)_n}{n!} \Gamma{\left(\frac{|\ell|}{2}+1\right)}.
\end{equation}
After applying further simplifications to the above equation, it leads to Eq. (\ref{eq:OAM_generation}).

\section{Derivation of Eq. (\ref{eq:LG_amplitudes})}
\label{App: App_B}
In this section, $\text{LG}_{\ell_i,p_i}$ after passing through an SPP with topological charge ${\ell_0}$ is decomposed into LG modes to determine the corresponding amplitudes. Using Eq. \eqref{eq:A3}, the amplitude of LG modes decomposition is written as

\begin{equation}
\begin{split}
C_{\ell_i-\ell_0,n} =& \sqrt{\frac{n!}{(|\ell_i - \ell_0|+n)!}}\sqrt{\frac{p_i!}{(|\ell_i|+p_i)!}} \exp{\left[j(|\ell_i| - |\ell_i-\ell_0|)+2(p_i-n))\zeta(z)\right]} \\
&\int_0^\infty \exp({-\psi}) \psi^{\left(\frac{|\ell_i|}{2}+\frac{|\ell_i-\ell_0|}{2}\right)} L_n^{|\ell_i-\ell_0|}(\psi) L_{p_i}^{|\ell_i|}(\psi) d\psi.
\end{split}
\label{eq:B1}
\end{equation}
For the calculation of the integral in the above equation, one of the Laguerre functions should written in explicit form. The explicit form of the Laguerre function is described as \cite{arfken2011mathematical}
\begin{equation}
L_n^k(\psi) = \sum_{m=0}^n (-1)^m \frac{(n+k)!}{(n-m)!(k+m)!m!}\psi^m.
\end{equation}
If the explicit form of the Laguerre function is used, the last integral in Eq. \eqref{eq:B1} can be calculated using Eq. \eqref{eq:A4}. After applying some simplifications, it leads to Eq. \eqref{eq:LG_amplitudes}.

\section{Derivation of Eq. (\ref{eq:P_eff1})}
\label{App: App_C}
In order to calculate the efficiency, first, rewrite Eq. (\ref{eq:I_eff}) in terms of magnitude and phase

\begin{equation}
I_{\text{eff}}(\rho,z) = \left|\sum_n \left|C_{0,n}\right| \left|\text{LG}_{0,n}\right| \exp{\left[-jk \left(z +\frac{\rho^2}{2R(z)} \right) + j(|\ell_i|+2p_i+1))\zeta (z)\right]} \right|^2
\end{equation}
In the above equation, the phase term is independent of $n$, so it would be enough to consider only the magnitude of the argument in the summation, which leads to
\begin{equation}
I_{\text{eff}}(\rho,z) = \left|\sum_n \left|C_{0,n}\right| \left|\text{LG}_{0,n}\right| \right|^2.
\end{equation}
The above summation includes mutual terms involving the multiplication of LG modes with different radial indices but identical azimuthal indices. While these LG modes are orthogonal when integrated over the entire plane $(\rho,\phi)$, the integration over $\rho$, here is limited to the circular aperture radius $R$. However, since the power of the LG modes is concentrated near the center of the plane, it can be assumed that the integration over the mutual terms is negligible. Consequently, the efficiency can be calculated as

\begin{equation}
P_{\text{eff}} \approx \sum_n \left|C_{0,n}\right|^2 \int_0^{2\pi} \int_0^R \left|\text{LG}_{0,n}(\rho,z)\right|^2 \rho d\rho d\phi.
\end{equation}
By applying some simplifications and substituting $\psi = 2\rho^2/W^2(z)$, the following expression is obtained

\begin{equation}
P_{\text{eff}} \approx \sum_n \left|C_{0,n}\right|^2 A_{0,n}^2 \frac{\pi}{2} \int_0^{\frac{2R^2}{W^2(z)}} \left[L_0^n(\psi)\right]^2 \exp(-\psi) d\psi.
\end{equation}
For the calculation of the integral in the above equation, the explicit forms of both generalized Laguerre polynomials must be used. Thus,

\begin{equation}
P_{\text{eff}} \approx \frac{\pi}{2} \sum_n \left|C_{0,n}\right|^2 A_{0,n}^2 \sum_{m=0}^n a_m \sum_{k=0}^n a_k \int_0^{\frac{2R^2}{W^2(z)}} \psi^{m+k} \exp(-\psi) d\psi.
\end{equation}
The integral in the above equation can be evaluated using the incomplete gamma function, defined as follows

\begin{equation}
\int_0^a \psi^n \exp(-\psi) d\psi= \gamma(n+1,a).
\end{equation}
By utilizing the incomplete gamma function and applying some simplifications, Eq. \eqref{eq:P_eff1} would be obtained. Furthermore, crosstalk in the first scenario, as defined in Eq. \eqref{eq:P_XT1}, can be derived in a similar manner.

\section{Derivation of Eq. (\ref{eq:P_eff2})}
\label{App: App_D}
In this section, the efficiency in the second scenario is calculated. First, it is assumed that the $\text{OAM}_{\ell_1}$ beam passes through an SPP with an inverse topological charge, after propagating from the transmitter to the receiver over a distance of $d_1$. It can be described as
\begin{equation}
U_{\ell_1}^{\text{det}}(\rho,\phi,z) = \sum_n C_{\ell_1,n} \text{LG}_{\ell_1,n}(\rho,\phi,z+d_1)\exp(j \ell_1\phi),
\end{equation}
where $U_{\ell_1}^{\text{det}}(\rho,\phi,z)$ represents $\text{OAM}_{\ell_1}$ after passing through $\text{SPP}_{-\ell_1}$. The received beam is then decomposed into the basis functions as follows

\begin{equation}
U_{\ell_1}^{\text{det}}(\rho,\phi,z) = \sum_p E_{0,p} \text{LG}_{0,p}(\rho,\phi,z+d_1),
\label{eq:32}
\end{equation}
where $E_{0,p}$ can be derived like Appendix \ref{App: App_A} and \ref{App: App_B} with a little more complexity
\begin{equation}
\begin{split}
E_{0,p} =& \sum_n \exp{\left[j(|\ell_1| + 2(n-p))\zeta(z+d_1) -j(|\ell_1| + 2n)\zeta(z)\right]} \frac{|\ell_1|}{2n!(|\ell_1|+n)!} \Gamma\left(\frac{|\ell_1|}{2}+n\right) \\ 
&\sum_{m=0}^p \frac{(-1)^m p!}{(p-m)!(m!)^2} \left(\frac{|\ell_1|}{2}-m\right)_n \Gamma{\left(\frac{|\ell_1|}{2}+m+1\right)}.
\end{split}
\end{equation}
Hence, the intensity of the received beam can be described as

\begin{equation}
I_{\text{eff}} = \left|\sum_p E_{0,p} \text{LG}_{0,p}(\rho,\phi,z+d_1)\right|^2.
\label{eq:D4}
\end{equation}

To eliminate the phase effect in the above equation, it is assumed that $\zeta(z+d_1) = \zeta(z) + \pi$. This is a reasonable assumption, indicating that the beam is before the beam waist at the transmitter and after the beam waist at the receiver. Finally, similar to the steps in Appendix \ref{App: App_C}, the efficiency can be calculated using the intensity derived in Eq. \eqref{eq:D4}. Additionally, the crosstalk in the second scenario, as defined in Eq. \eqref{eq:P_XT2}, can be derived in a similar manner.


\end{document}